\documentclass[amsmath,amssymb
   ,final     ,numberedheadings        
  ]
 {aipproc}

\layoutstyle{6x9}
\usepackage{bm}

\newcommand{\ads}{\ensuremath{\mathrm{AdS}_5\,}}
\newcommand{\ZZ}{\ensuremath{\mathbf{Z}}}
\newcommand{\kappafive}{\kappa_5}
\newcommand{\kappareduct}{\kappa_4}
\newcommand{\tension}{\sigma}

\newcommand{\bb}{{\scriptscriptstyle{\bullet}}}
\newcommand{\p}{\partial}
\newcommand{\ud}{\mathrm{d}}
\newcommand{\Hb}{h_{\bb}}

\newcommand{\bk}{\mathbf{k}}
\newcommand{\bx}{\mathbf{x}}
\newcommand{\vv}{{v}}

\newcommand{\de}{\delta}
\newcommand{\ga}{\gamma}

\newcommand{\dd}{\partial}
\newcommand{\ra}{\rightarrow}
\newcommand{\al}{\alpha}
\newcommand{\om}{\omega}
\newcommand{\lsim}{\,\raisebox{-0.6ex}{$\buildrel < \over \sim$}\,}
\newcommand{\gsim}{\,\raisebox{-0.6ex}{$\buildrel > \over \sim$}\,}
\newcommand{\be}{\begin{equation}}
\newcommand{\ee}{\end{equation}}
\newcommand{\ben}{\begin{equation*}}
\newcommand{\een}{\end{equation*}}
\newcommand{\bea}{\begin{eqnarray}}
\newcommand{\eea}{\end{eqnarray}}
\newcommand{\bean}{\begin{eqnarray*}}
\newcommand{\eean}{\end{eqnarray*}}

\begin{document}

\title{Graviton production in brane worlds by the dynamical Casimir effect}


\author{Ruth Durrer}{
  address={D\'epartement de Physique Th\'eorique, Universit\'e de Gen\`eve,
  24, Quai E. Ansermet, 1211 Gen\`eve 4, Switzerland},
  email={ruth.durrer@unige.ch}
}

\author{Marcus Ruser}{
  address={D\'epartement de Physique Th\'eorique, Universit\'e de Gen\`eve,
  24, Quai E. Ansermet, 1211 Gen\`eve 4, Switzerland},
   email={marcus.ruser@gmail.com}
}

\author{Marc Vonlanthen}{
  address={D\'epartement de Physique Th\'eorique, Universit\'e de Gen\`eve,
  24, Quai E. Ansermet, 1211 Gen\`eve 4, Switzerland} ,
 email={marc.vonlanthen@unige.ch}
}

\author{Peter Wittwer}{
  address={D\'epartement de Physique Th\'eorique, Universit\'e de Gen\`eve,
  24, Quai E. Ansermet, 1211 Gen\`eve 4, Switzerland} ,
 email={peter.wittwer@unige.ch}
}
\begin{abstract}
If our Universe is a $3+1$ brane in a warped $4+1$ dimensional bulk so that 
its expansion can be understood as the motion of the brane in the bulk, the
time dependence of the boundary conditions for arbitrary bulk fields can lead
to particle creation via the dynamical Casimir effect. In this talk I report results
for the simplest such scenario, when the only particle in the bulk is the graviton
and the bulk is the 5 dimensional anti-de Sitter spacetime.  
\end{abstract}

\maketitle


\section{Introduction}
The idea that our Universe be a $3+1$ dimensional membrane in a higher 
dimensional 'bulk' spacetime has opened new exciting prospects for 
cosmology, for reviews see~\cite{roy,mine}. In the simplest braneworlds 
motivated by string theory, the standard model particles are confined to 
the brane and only the graviton can propagate in the bulk. Of particular 
interest is the Randall-Sundrum (RS) model~\cite{RS1,RS2}, where the bulk 
is  5-dimensional anti-de Sitter space, AdS$_5$. If the so called RS fine tuning 
condition is satisfied, it can be shown that gravity on the brane 
'looks 4-dimensional' at low energies.

Within this model, cosmological evolution can be interpreted as the 
motion of the physical brane, i.e. our Universe, through the 5d bulk.
Such a time-dependent boundary does in general lead to particle 
production via the dynamical Casimir effect~\cite{bordag}.

Of course one can always choose coordinates with respect to which the brane 
is at rest, e.g. Gaussian normal coordinates. But then usually (except in 
the case of de Sitter expansion on the brane~\cite{ruba}), the perturbation 
equation describing the evolution of gravitons is not separable and can be 
treated only with numerical simulations~\cite{jap1,kazu,seahra}. 
Furthermore, in a time-dependent bulk 
a mode decomposition is in general ambiguous and one cannot split 
the field in a zero mode and Kaluza-Klein (KK) modes in a unique way.

\begin{figure}[ht]
\centering
\includegraphics[height=7cm]{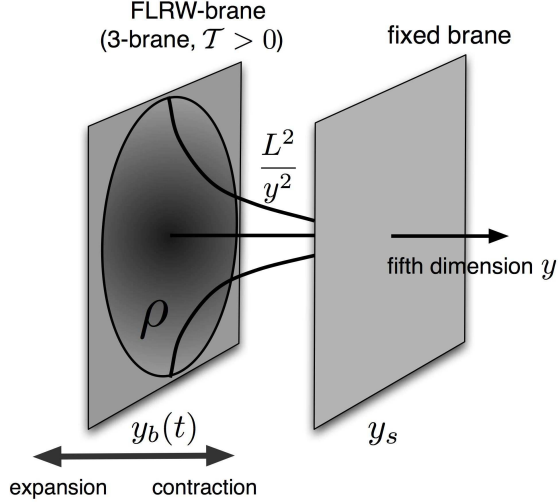}
\caption{\label{f:1} Two branes in an AdS$_5$ spacetime. The physical
  brane is to the left. While it is approaching the static brane its scale 
  factor is decreasing, the Universe is contracting, and when it moves away 
  from the static brane the Universe is expanding. The AdS curvature 
  radius $L$ (dashed line) 
  and value of the scale factor of the brane metric as function of the 
  extra dimension $y$  (light (blue) line) are also indicated.}
\end{figure}

Based on the picture of a moving brane in AdS$_5$, we have studied graviton 
production in an ekpyrotic type scenario~\cite{ekpy} where our Universe 
first approaches a second static brane. After a 'collision' the physical 
brane reverses direction and moves away from the static brane, see 
Fig.~\ref{f:1}. For an observer 
on the brane, the first phase corresponds to a contracting Universe and the 
collision represents the 'Big Bang' after which the Universe starts expanding.

Here I report on the results which we  have obtained in our previous 
papers~\cite{DR,RD,RDVW}. We have found that the energy density of KK 
gravitons in AdS$_5$ scales like stiff matter, $\rho_{\mathrm{KK}} \propto a^{-6}$, 
here $a$ denotes the  scale factor defined in Eq.~(\ref{eq:branemetric}). 
Therefore, KK  gravitons in AdS$_5$ cannot represent the dark matter in 
the Universe. This finding is in contrast with the results of Ref.~\cite{lang} and
we comment on this below. We have also found that in the early Universe the back reaction 
from KK gravitons on the bulk geometry is likely to be important. 

Finally, we 
have derived a limit for the maximal brane velocity, the bounce velocity, 
$v_b\lsim 0.2$ in order not to over-produce zero-mode (i.e. 4d) gravitons, 
the energy density of which is constrained by the nucleosynthesis bound. We have
calculated the spectra of both, the zero-mode and the KK gravitons.
In Refs.~\cite{DR,RD} we have, however, neglected a term linear in the 
brane velocity $v$ in the boundary conditions. In our latest work,
Ref.~\cite{RDVW} we derived a method which includes this term and 
allows to treat the problem without any low  velocity approximation. We have shown that 
the low velocity results previously obtained are not modified. 

The remainder of this paper is organized as follows. In the next section we
present the basic equations for the evolution of tensor perturbations (gravitons)
and we explain why it is not straight forward to include the velocity term 
of the boundary condition. In Section~3 we quantize the system. In 
Section~4 we discuss our results and in Section~5 we conclude.

\section{A moving brane in $\ads$}
%
\subsection{The background}
In Poincar\'e coordinates $(x^A)=(t, {\bf x}, y)$ with ${\bf x} = (x^1, x^2, x^3)$ and $A=0,...,4$, 
the AdS$_5$ (bulk) metric is given by
\begin{equation}\label{e:bulk-metric}
 ds^2
 = g_{AB} d x^{A} d  x^{B}
 = \frac{L^2}{y^2} \left[-d t^2 + \delta_{ij} d x^i d x^j + dy^2\right]~,
\end{equation}
where $i,j=1,2,3$ and $L$ is the AdS$_5$ curvature radius  which is related
to the bulk cosmological constant by the 5d Einstein equation, 
$-\Lambda=6/L^2$. The physical brane representing our (spatially flat) 
Universe is located at some time dependent
position $y=y_b(t)$ in the bulk, and the metric induced on the brane 
is the Friedman-Robertson-Walker metric,
\begin{eqnarray}
 ds^2 = a^2(\eta)\left[ -\ud\eta^2 +\delta_{ij}d x^id x^j\right]~,
 \label{eq:branemetric}
\end{eqnarray}
with scale factor $a(\eta)$ which is given by the brane position,
\begin{equation}
a(\eta)=\frac{L}{y_b(t)}~.
\label{e:a of y}
\end{equation}
The conformal time  $\eta$ of an observer on the brane,
 is related to the bulk time $t$ via
\begin{equation}
 d\eta = \sqrt{1- v^2}dt \equiv \ga^{-1}dt~.
\end{equation}
Here we have introduced the brane velocity
\be\label{e:6}
 v \equiv \frac{dy_b}{dt} = -\frac{LH}{\sqrt{1+L^2H^2}}\quad \mbox{
 and }~  \ga  = \frac{1}{\sqrt{1-v^2}} ~.
\ee
$H$ is the usual Hubble parameter,
\begin{equation}
   H \equiv \frac{1}{a^2}\frac{\partial a}{\partial \eta} \equiv 
   a^{-1}{\cal H}= -L^{-1}\ga v~.
\end{equation}
The brane dynamics, as a result of the second junction condition, is determined 
by the modified Friedmann equation~\cite{roy}
\begin{equation}
H^2 = \frac{\kappa_4 \rho}{3} \left( 1+ \frac{\rho}{2{\sigma}}\right)
\end{equation}
where ${\sigma}$ is the brane tension, $\rho$ the energy density on the 
brane, and we assume the RS fine tuning condition~\cite{RS1}
 \begin{equation}
 \frac{\kappafive^2 \tension^2}{12} = \frac{3}{L^2}~,
 \label{e:fine}
\qquad \mbox{ and } \qquad
 \kappareduct \equiv 8\pi G_4 \equiv \frac{\kappafive^2 \tension}{6}~.
\end{equation}
 We define the string and Planck scales by 
\begin{equation}
\kappa_5  =\frac{1}{M_5^3} = L_s^3~, \qquad
\kappa_4 = \frac{1}{M_{\rm Pl}^2}= L_{\rm Pl}^2~.
\label{e:string and Pl scale}
\end{equation}
Note that the RS fine-tuning condition is equivalent to
\begin{equation}
\kappa_5=\kappa_4\,L~ \mbox{ or }~~ \frac{L_s}{L}=\frac{L_{\rm Pl}^2}{L_s^2}. 
\label{e:RS fine tuning 0}
\end{equation}
%
\subsection{Tensor perturbations}
%
We now consider 3d tensor perturbations
$h_{ij}(t,{\bf x},y)$
of the spatial three-dimensional geometry on this background. 
The perturbed bulk metric reads 
\be
 ds^2 = \frac{L^2}{y^2}
 \left[-dt^2+(\delta_{ij}+2h_{ij})dx^i dx^j+d y^2 \right]~.
\ee
Tensor modes satisfy the traceless and transverse conditions, $h_i^i =
\p_ih^i_j = 0$. 
These conditions imply that $h_{ij}$ has only two independent degrees 
of freedom, the two polarization states $\bullet=\times,+$.
We decompose $h_{ij}$ into spatial Fourier modes,
\begin{equation}
 h_{ij}(t,{\bx},y)
 = \int \frac{d^3k}{(2\pi)^{3/2}} \sum_{\bb=+,\times}
  e^{i\bk\cdot{\bx}}e_{ij}^{\bb}({\bf k})\Hb(t,y;{\bf k})~,
\label{e:h fourier decomposition}
\end{equation}
where $e_{ij}^{\bb}({\bf k})$ are unitary constant transverse-traceless
polarization tensors which form a basis of the two polarization
states $\bullet = \times,+$. Since we assume parity 
symmetry, we shall neglect in the
following the distinction between the two graviton polarizations and
consider only one of them. We then have to multiply the final results
for e.g. particle number or energy density by a factor of two to account 
for both polarizations.

The perturbed Einstein equations and the second junction condition 
lead to the following boundary value problem 
\begin{equation}
 \left[\p_t^2 +k^2 -\p_y^2 + \frac{3}{y}\p_y \right] h(t,y;{\bf k}) = 0~~{\rm in}~{\rm the}~{\rm bulk,}
 \quad k^2 = |{\bf k}|^2\,,
\label{e:T-bulk-eq}
\end{equation}
and
\begin{equation}
        \left . \ga\left(\vv\dd_t +\dd_y\right)h \right|_{y_b(t)}=
         0~~{\rm on}~{\rm the}~{\rm brane}~.
\label{e:T-JC-simple}
\end{equation}
 We introduce also a second, static brane at position $y_s$, 
which requires the additional boundary condition
\begin{equation}
\left . \dd_yh \right|_{y_s}=0~~\mbox{on the static brane}~.
\label{e:T-JC-stat}
\end{equation}

Eq.~(\ref{e:T-bulk-eq}) is the Klein-Gordon equation for a minimally coupled
massless mode in $\ads$,
i.e. the operator acting on $h$ is just the Klein-Gordon operator 
\begin{equation}
          \raisebox{3pt}{\fbox{ ~}} = \frac{1}{\sqrt{-g}} \partial_A 
          \left[ \sqrt{-g} g^{AB} \partial_B\right]~.
\label{e:KG operator}
\end{equation}

Equation (\ref{e:T-JC-simple})  is the time-dependent boundary 
condition (BC) coming from the fact that the moving brane acts like a "moving 
mirror" for the gravitational perturbations. Only in the rest-frame of the brane
do we have pure Neumann BC. In a generic frame we have the 
Lorentz transformed BC which contains a velocity term $v\dd_t$.

We assume that the brane is filled with a perfect fluid
such that there are no anisotropic stress perturbations in the 
brane energy momentum tensor, i.e. there is no coupling of gravitational waves 
to matter. If this were the case, the r.h.s. of Eq.~(\ref{e:T-JC-simple}) would 
not be zero but a term coupling $h_{ij}$ to the matter on the brane,
see Eq.~(2.25) of \cite{RD}. 

For the tensor perturbations the gravitational action up to
second order in the perturbations reads
\be
{\cal S}_h = 4\,\frac{L^3}{2\kappa_5} \int dt\! \int d^3k 
\!\int_{y_b(t)}^{y_s} \frac{dy}{y^3}  \Big[|\partial_t h|^2 
- |\partial_y h|^2  -k^2|h|^2 \Big]~.
\label{e:action h}
\ee
One factor of two in the action is due to ${\ZZ}_2$ symmetry while
a second factor comes from the two polarizations.
%
\subsection{Dynamical Casimir effect approach}
%
 The wave equation (\ref{e:T-bulk-eq}) itself has no time dependence and
simply describes the propagation of free modes. It is
the time dependence of the BC (\ref{e:T-JC-simple}) that sources the
non-trivial time-evolution of the perturbations. 
As it is well known, such a system of a wave equation and time-dependent
BC lead, within a quantum mechanical formulation, to particle production
from vacuum fluctuations. 
In the context of the photon field perturbed by a moving mirror this 
goes under the name ``dynamical Casimir 
effect''~\cite{bordag}.

In \cite{RD} we have extended a formalism which has been successfully 
employed for the numerical investigation of photon production in dynamical 
cavities \cite{Ruser:2004,Ruser:2005xg,Ruser:2006xg} to the RS braneworld 
scenario. We have studied graviton production by a moving brane,  which we 
call dynamical Casimir effect for gravitons, for a bouncing braneworld scenario.

However, in order to solve the problem, we have neglected the velocity term 
in Eq.~(\ref{e:T-JC-simple}). The ansatz
$$ h= \sum_\al a_\al(t)e^{-i\om_\al t}\phi_\al(t,y) + {\rm h.c.}~, ~~ 
  \om_\al^2 =k^2 +m_\al(t)^2$$
then leads to a Sturm--Liouville problem for the instantaneous 
eigenfunctions $\phi_\al(t,y)$  which satisfy 
\be\label{e:mode}
\left(-\dd_y^2+ \frac{3}{y}\dd_y\right)\phi_\al = m_\al^2\phi_\al~. 
\ee
The  solutions of (\ref{e:mode}) are
\bea
\phi_0(t) &=& \frac{y_s y_b(t)}{\sqrt{y_s^2 - y_b^2(t)}}\, ,
\label{zero mode phi}\\
\phi_n(t,y) &=& N_n (t) y^2C_2(m_n(t),y_b(t),y)  \qquad ~ \mbox{ with}
\nonumber \\   \hspace*{-2mm} \label{massive mode phi}
C_\nu(m,x,y) &=& Y_1(m x) J_\nu(my)\! -\! J_1(m x) Y_\nu(m y)\, .
\eea
The function $\phi_0$ is the zero mode which corresponds to the ordinary
$(3+1)$d graviton on the brane while the $\phi_n$ are the KK modes. 
The masses $m_n$ are determined by the boundary condition at the static 
brane, see, e.g.~\cite{CDR} for more details.
Since $\phi_\al$ satisfies Neumann boundary conditions, 
we know that the solutions $(\phi_\al)_\al$ form a complete orthonormal set 
of functions on the interval $[y_b(t),y_s]$ normalized by the scalar product
$$ \left(\phi_\al,\phi_\beta\right) \equiv 2\int_{y_b(t)}^{y_s}\frac{dy}{y^3}\phi_\al\phi_\beta 
   =\de_{\al\beta}~.$$
Therefore, any general solution which satisfies Neumann BC can be expanded 
in these instantaneous eigenfunctions.
If we add the term $v\dd_t$ to the boundary condition this feature is lost, and
we can no longer expect to find a complete set of instantaneous eigenfunctions.

However, since the entire effect disappears when the velocity tends to zero,
neglecting a term which is first order in the velocity seems not to be
consistent. This problem led us to search for another 
approach which is discussed in Ref.~\cite{RDVW} where we transform to a
coordinate system where the velocity term disappears identically. There also show that for
low velocities $v<0.3$, say the corrections obtained with this consistent treatment are 
below a few percent. We therefore ignore it in the following.

\section{Quantization}

\subsection{Equation of motion}
The gravitational wave amplitude $h(t,y;{\bf k})$ subject to Neumann 
boundary conditions  can be expanded as
\begin{equation}
h (t,y;{\bf k}) = \sqrt{\frac{\kappa_5}{L^3}}\sum_{\alpha = 0}^\infty
q_{\alpha,{\bf k}}(t)\phi_\alpha(t,y)~.
\label{e: mode expansion}
\end{equation}
The coefficients  $q_{\alpha,{\bf k}}(t)$
are canonical variables describing the time evolution of 
the perturbations and the factor $\sqrt{\kappa_5/L^3}$ 
has been introduced in order to render the 
$q_{\alpha,{\bf k}}$'s canonically normalized.
For $h(t,y,\bx)$  to be real, we have to impose
the following reality condition on the canonical variables, 
\begin{equation}
q_{\alpha,{\bf k}}^* = q_{\alpha,{\bf -k}}\, .
\label{e:reality for q}
\end{equation}

One could now insert the expansion (\ref{e: mode expansion}) 
into the wave equation (\ref{e:T-bulk-eq}), 
multiply it by $\phi_\beta(t,y)$ and integrate out the 
$y-$dependence by using the orthonormality 
to derive the equations of motion for the variables
$q_{\alpha, {\bf k}}$. However, as we explain in Refs.~\cite{RD,RDVW}, 
 a Neumann boundary condition at a moving brane is not compatible with a free 
wave equation. The only consistent way to implement Neumann boundary 
conditions is therefore
to consider the action (\ref{e:action h}) of the perturbations
as the starting point to derive the equations of
motion for $q_{\alpha,{\bf k}}$.  
Inserting (\ref{e: mode expansion}) into (\ref{e:action h}) 
leads to the action 
\bea
\hspace*{-0.7cm}{\cal S} &=& \frac{1}{2} \int dt \int d^3k \Big\{ 
\sum_\alpha \left[ |\dot{q}_{\alpha,{\bf k}}|^2 -
\omega_{\alpha,k}^2|q_{\alpha,{\bf k}}|^2\right]  + \nonumber \\
 && \qquad \qquad\qquad \sum_{\alpha\beta}\big[M_{\alpha\beta} \left(q_{\alpha,{\bf k}}
\dot{q}_{\beta,{\bf -k}} + q_{\alpha,{\bf -k}}
\dot{q}_{\beta,{\bf k}}\right)  
+ N_{\alpha\beta}
q_{\alpha,{\bf k}}q_{\beta,{\bf -k}}\big]\Big\}~.  \label{e:action}
\eea
We have introduced the time-dependent frequency of a graviton mode
\begin{equation}
\om_{\al, k}^2 = \sqrt{ k^2 +m_\al^2}\;,
\end{equation}
and the time-dependent coupling matrices 
\begin{eqnarray}
M_{\alpha\beta} &=& (\partial_t \phi_\alpha, \phi_\beta)~,\\
N_{\alpha\beta} &=& (\partial_t \phi_\alpha, \partial_t
\phi_\beta)=\sum_\gamma M_{\alpha\gamma}M_{\beta\gamma} = (MM^T)_{\alpha\beta}\,,
\end{eqnarray}
which are given explicitely in Ref.~\cite{RD} (see also \cite{CDR}).
The equations of motion for the canonical variables  
are the Euler--Lagrange equations from the action (\ref{e:action}), 
\be
\ddot{q}_{\alpha,{\bf k}} + \omega_{\alpha, k}^2 q_{\alpha,{\bf k}} + \sum_\beta
\left[M_{\beta\alpha} - M_{\alpha\beta}\right]\dot{q}_{\beta,{\bf k}}
+\sum_\beta\left[\dot{M}_{\alpha\beta} -
  N_{\alpha\beta}\right]q_{\beta,{\bf k}} = 0 ~.
\label{deq for q}
\ee 

The motion of the brane through the bulk, i.e. the expansion 
of the universe, is encoded in the time-dependent coupling matrices 
$M_{\alpha\beta}$ and $ N_{\alpha\beta}$.
These mode couplings are caused by the time-dependent boundary condition
$\partial_y h_\bullet(t,y)|_{y_b} = 0$ which forces the eigenfunctions
$\phi_\alpha(t,y)$ to be explicitly time-dependent.
In addition, the frequency of the KK modes $\omega_{\alpha,k}$
is also time-dependent since the distance between the two branes 
changes when the brane is in motion.
Both time dependencies can lead to the amplification 
of tensor perturbations and, within a quantum treatment
which is developed below, to graviton production from vacuum.

Because of translational invariance with respect to the 
directions parallel to the brane, modes with different 
${\bf k}$ do not couple in (\ref{deq for q}).
The three-momentum ${\bf k}$ enters the equation of motion 
for the perturbation only via the frequency $\omega_{\alpha,k}$.
Equation (\ref{deq for q}) is similar to the equation
describing the time evolution of electromagnetic field modes 
within a three-dimensional dynamical cavity \cite{Ruser:2005xg} 
and may effectively be described by a massive scalar field 
on a time-dependent interval \cite{Ruser:2006xg}.
For the electromagnetic field, the dynamics of the 
cavity, or more precisely the motion of one of its walls,
leads to photon creation from vacuum fluctuations.
This phenomenon is usually referred to as dynamical 
Casimir effect.
Inspired by this, we call the production of gravitons
by the moving brane the {\it dynamical Casimir effect for 
gravitons}.

\subsection{Quantization}
Asymptotically, i.e. for $t\rightarrow \pm \infty$,
the physical brane approaches the Cauchy horizon
($y_b\rightarrow 0$), moving very slowly.
Then, the coupling matrices vanish and the KK masses
become constant,
\begin{equation}
\lim_{t\ra \pm \infty}M_{\alpha\beta}(t)=0\;\;,\;\;
\lim_{t\ra \pm \infty} m_\alpha(t) = {\rm const.}
\;\;\forall \alpha,\beta
\;\;.
\end{equation}
In this limit, the system~(\ref{deq for q}) reduces to an infinite set of uncoupled
harmonic oscillators.
This allows to introduce an unambiguous and meaningful particle
concept, i.e. the notion of (massive) gravitons.

Canonical quantization of the gravity wave amplitude is performed by replacing the
canonical variables $q_{\alpha, {\bf k}}$ by the corresponding operators
$\hat{q}_{\alpha, {\bf k}}$
\begin{equation}
\hat{h}(t,y;{\bf k}) = \sqrt{\frac{\kappa_5}{L^3}}\sum_\alpha
\hat{q}_{\alpha,{\bf k}}(t) \phi_\alpha(t,y)~.
\label{e:expansion of h bullet in q}
\end{equation}
Adopting the Heisenberg picture to describe the quantum
time evolution, it follows that  $\hat{q}_{\alpha, {\bf k}}$
satisfies the same equation (\ref{deq for q}) as the canonical variable
$q_{\alpha, {\bf k}}$.

Under the assumptions outlined above, the operator $\hat{q}_{\alpha, {\bf k}}$
can be written for very early times, $t < t_{\rm in}$, as
\be\label{e:initial q}
\hat{q}_{\alpha, {\bf k}}(t < t_{\rm in}) = 
\frac{1}{\sqrt{2\omega_{\alpha, k}^{\rm in}}}
\left[ \hat{a}^{\rm in}_{\alpha, {\bf k}} e^{-i\,\omega_{\alpha,k}^{\rm in}\,t}
+\hat{a}^{{\rm in} \dagger}_{\alpha, -{\bf k}} e^{i\,\omega_{\alpha,k}^{\rm in}\,t}
\right]\,,
\ee
where we have introduced the reference frequency
\begin{equation}
\omega_{\alpha,k}^{\rm in}  \equiv  \omega_{\alpha,k}(t < t_{\rm in})~.
\end{equation}

This expansion ensures that Eq.~(\ref{e:reality for q}) is satisfied.
The set of annihilation and creation operators
$\{\hat{a}^{\rm in}_{\alpha, {\bf k}}$,
$\hat{a}^{{\rm in} \dagger}_{\alpha, {\bf k}} \}$
corresponding to the notion of gravitons for
$t < t_{\rm in}$ is subject to the usual commutation relations
\begin{eqnarray}
\left[\hat{a}^{\rm in}_{\alpha, {\bf k}},
\hat{a}^{{\rm in}\dagger}_{\alpha', {\bf k}'}\right]
&=&\delta_{\alpha\alpha'} \delta^{(3)}({\bf k} - {\bf k'})\;,\\
\Big[\hat{a}^{\rm in}_{\alpha, {\bf k}},
\hat{a}^{\rm in}_{\alpha', {\bf k'}}\Big]
&=&
\left[\hat{a}^{{\rm in} \dagger}_{\alpha, {\bf k}},
\hat{a}^{{\rm in} \dagger}_{\alpha', {\bf k'}}\right]
=0.
\end{eqnarray}
For very late times, $t > t_{\rm out}$, i.e. after the motion of the brane has
ceased, the operator
$\hat{q}_{\alpha, {\bf k}}$  can be expanded in a similar manner,
\bea\label{e:final q}
\hat{q}_{\alpha, {\bf k}}(t > t_{\rm out}) = 
\frac{1}{\sqrt{2\omega_{\alpha, k}^{\rm out}}}
\left[ \hat{a}^{\rm out}_{\alpha, {\bf k}} e^{-i\,\omega_{\alpha,k}^{\rm out}\,t}
+\hat{a}^{{\rm out}\,\dagger}_{\alpha, -{\bf k}} e^{i\,\omega_{\alpha,k}^{\rm out}\,t}
\right] 
\eea
with final state frequency
\begin{equation}
\omega_{\alpha,k}^{\rm out}  \equiv  \omega_{\alpha,k}(t > t_{\rm out})~.
\end{equation}
The annihilation and creation operators
$\{\hat{a}^{\rm out}_{\alpha, {\bf k}} ,
\hat{a}^{{\rm out} \, \dagger}_{\alpha, {\bf k}} \}$
correspond to a meaningful definition of final state gravitons
(they are associated with positive and negative
frequency solutions for $t\ge t_{\rm out}$)
and satisfy the same commutation relations as the initial state operators\footnote{Of course the 
brane never really stops moving, but before a certain time $t_{\rm in}$ and after a certain time
$ t_{\rm out} $ the motion is so slow that no particle production takes place. We have chosen these times sufficiently early (rsp. late) so that the numerical results are independent of their choice.}.
 
Initial $|0,{\rm in}\rangle \equiv |0,t <  t_{\rm in} \rangle$ and final
$|0,{\rm out}\rangle \equiv |0,t >  t_{\rm out} \rangle$ vacuum
states are uniquely defined via 
\footnote{Note that the notations $|0,t <  t_{\rm in} \rangle$ and $|0,t >  t_{\rm out} \rangle$
do not mean that the states are time-dependent; 
states do not evolve in the Heisenberg picture.}
\begin{equation}
\hat{a}^{\rm in}_{\alpha,{\bf k}}|0,{\rm in}\rangle = 0 \;,\;
\hat{a}^{\rm out}_{\alpha,{\bf k}} |0,{\rm out}\rangle = 0 \;,\;\;
\forall \;\alpha,\;{\bf k}~.
\label{vacuum definitions}
\end{equation}
The operators counting the number of particles defined with respect to the initial
and final vacuum state, respectively, are
\begin{equation}
\hat{N}^{\rm in}_{\alpha, {\bf k}} =  \hat{a}^{{\rm in}\,\dagger}_{\alpha,{\bf k}}
\hat{a}^{\rm in}_{\alpha,{\bf k}}\;,\;\;
\hat{N}^{\rm out}_{\alpha, {\bf k}} =  \hat{a}^{{\rm out} \,\dagger}_{\alpha,{\bf k}}
\hat{a}^{\rm out}_{\alpha,{\bf k}}~.
\end{equation}
The number of gravitons created during the motion of the brane
for each momentum ${\bf k}$ and quantum number $\alpha$ 
is given by the expectation value of the number operator
$\hat{N}^{\rm out}_{\alpha,{\bf k}}$
of final-state gravitons with respect to the initial
vacuum state $| 0,{\rm in} \rangle$:
\begin{equation}
{\cal N}^{\rm out}_{\alpha,{\bf k}} = \langle 0,{\rm in}|
\hat{N}^{\rm out}_{\alpha,{\bf k}}|0,{\rm in}\rangle.
\label{e:graviton number definition}
\end{equation}
If the brane undergoes a non-trivial dynamics between
$t_{\rm in} < t < t_{\rm out}$ we have
$\hat{a}^{\rm out}_{\alpha,{\bf k}}|0,{\rm in}\rangle \neq 0$ 
in general, i.e. graviton production from vacuum fluctuations 
takes place.

\section{Results}
\subsection{Energy density}
For a usual four-dimensional tensor perturbation $h_{\mu\nu}$
on a background metric $g_{\mu\nu}$ an associated effective
energy momentum tensor can be defined unambiguously by
\begin{equation}
T_{\mu\nu} = \frac{1}{\kappa_4} \langle
h_{\alpha\beta\|\mu}h^{\alpha\beta}_{\;\;\;\;\|\nu}\rangle ~,
\end{equation}
where the bracket stands for averaging over 
several periods of the wave and ``$\|$'' denotes
the covariant derivative with respect to the unperturbed
background metric.
The energy density of gravity waves is the
$00$-component of the effective energy momentum tensor.
We shall use the same effective energy momentum tensor to calculate
the energy density corresponding to the 
four-dimensional spin-2 graviton component of the five-dimensional tensor
perturbation on the brane, i.e. for
the perturbation $h_{ij}(t,{\bf x}, y_b)$. 
For this it is important to remember that in our low energy
approach, and in particular at very late times for which we 
want to calculate the energy density, the conformal 
time $\eta$ on the brane is identical 
to the conformal bulk time $t$. 
The energy density of four-dimensional spin-2 gravitons 
on the brane produced during the brane 
motion is then given by
\begin{equation}
\rho = \frac{1}{\kappa_4\,a^2} \left \langle \left \langle 0, {\rm in} |
\dot{\hat{h}}_{ij} (t, {\bf x}, y_b)\dot{\hat{h}}^{ij} (t, {\bf x}, y_b)
|0,{\rm in}\right \rangle \right \rangle.
\end{equation}
Here the outer bracket denotes averaging over several 
oscillations, which we embrace from the very beginning. 
The factor $1/a^2$ comes from the fact that an over-dot 
indicates the derivative with respect to conformal time $t\simeq \eta$.
The detailed calculation given in Ref.~\cite{RD} leads to
\begin{equation}
\rho = \frac{2}{a^4}\sum_{\alpha}
\int \frac{d^3k}{(2\pi)^{3}}
\omega_{\alpha,k}{\cal N}_{\alpha,k}(t){\cal Y}^2_{\alpha}(a)~
\label{energy density}
\end{equation}
where again ${\cal N}_{\alpha,k}(t)$ is the instantaneous 
particle number and ${\cal Y}_\alpha$ is related to value of the wave 
function on the brane by 
$$
{\cal Y}_\alpha(a) =\frac{a}{L}\phi_\alpha(t,y_b(t)) \,.
$$
The factor two reflects the two polarizations.
At late times, $t > t_{\rm out}$, after particle creation has ceased, 
the energy density is\begin{equation}
\rho = \frac{2}{a^4}\sum_{\alpha}
\int \frac{d^3k}{(2\pi)^{3}}
\omega_{\rm \alpha,k}^{\rm out} \;{\cal N}_{\alpha,\bk}^{\rm out}\;
{\cal Y}^2_{\alpha}(a).
\label{e:energy density late time out number}
\end{equation}
This expression looks at first sight very similar to a ``naive'' 
definition of energy
density as integration over momentum space and summation over all
quantum numbers $\alpha$ of the energy
$\omega_{\rm \alpha,k}^{\rm out} \;{\cal N}_{\alpha,\bk}^{\rm out}$
of created gravitons. 
However, the important difference is the appearance of the function
${\cal Y}^2_{\alpha}(a)$ which exhibits a different dependence on
the scale factor for the zero mode compared to the KK-modes.

Let us decompose the energy density into zero mode and
KK contributions
\begin{equation}
\rho = \rho_0 + \rho_{KK}.
\end{equation}
Evaluating ${\cal Y}_0(a)$ one then obtains 
for the energy density of the massless zero mode 
\begin{equation}\label{4.15}
\rho_0 = \frac{2}{a^4}\int \frac{d^3k}{(2\pi)^{3}}
\,k \,{\cal N}_{0,\bk}^{\rm out}~.
\end{equation}
This is the expected behavior;
the energy density of standard four-dimensional gravitons 
scales like radiation.
\\
In contrast, the energy density of the KK-modes 
at late times is found to be
\begin{equation}
\rho_{\rm KK} = \frac{L^2}{a^6}\frac{\pi^2}{2} \sum_{n=1}^\infty
\int \frac{d^3k}{(2\pi)^{3}}
\omega_{n,k}^{\rm out} \;{\cal N}_{n,\bk}^{\rm out}\,m_n^2 Y_1^2(m_ny_s),
\end{equation}
which decays like $1/a^6$. 
As the universe expands, the energy density
of massive gravitons on the brane is therefore rapidly diluted.
The total energy density of gravitational waves in our universe
at late times is dominated by the standard four-dimensional 
graviton (massless zero mode).
In the large mass limit, $m_n y_s \gg 1$, $n \gg 1$,
the KK-energy density can be approximated by
\begin{equation}
\rho_{{\rm KK}} \simeq \frac{\pi L^2}{2a^6y_s}
\sum_n
\int \frac{d^3k}{(2\pi)^{3}}
\;{\cal N}_{n,\bk}^{\rm out}\,\omega_{n,k}^{\rm out}\; m_n ~.
\label{late time large mass KK energy density}
\end{equation}
Due to the factor $m_n$ coming from the function 
${\cal Y}_n^2$, i.e. from the normalization of the functions $\phi_n(t,y)$, 
in order for the summation over the KK-tower to converge,
the number of produced gravitons ${\cal N}^{\rm out}_{n,\bk}$
has to decrease faster than $1/m_n^3$ for large masses and
not just faster than $1/m_n^2$ as one might naively expect.
%
\subsection{Escaping of massive gravitons and localization of gravity}
\label{ss:escape}
%
As we have shown, the energy density 
of the KK modes scales, at late times when particle 
production has ceased, with the expansion of the 
universe like
\begin{equation}
 {\rho}_{\rm KK} \propto1/a^6\,,
\label{e:KK scalings}
\end{equation}
i.e. it  decays by a factor $1/a^2$ faster than the 
corresponding expression for the zero mode graviton
and behaves effectively like stiff matter. 
Mathematically, this difference arises from the 
distinct behavior of the functions ${\cal Y}_0(a)$
and ${\cal Y}_n(a),~n\ge 1\,,$ and is a direct consequence of the warping of the 
fifth dimension which affects the normalization of the mode functions
$\phi_\alpha$. But what is the underlying physics?
As we shall discuss now, this scaling behavior 
for the KK particles has indeed a straight forward 
very appealing physical interpretation.

First, the mass $m_n$ is a comoving mass. 
The (instantaneous) 'comoving' frequency or energy 
of a KK graviton is $\omega_{n,k} =\sqrt{k^2+m_n^2}$, 
with comoving wave number $k$. 
The physical mass of a KK mode measured
by an observer on the brane with cosmic time $d\tau =adt$ is therefore
$m_n/a$, i.e. the KK masses are redshifted with the expansion of the
universe. This comes from the fact that $m_n$ is the wave number
corresponding to the $y$-direction with respect to the bulk time $t$
which corresponds to {\it conformal time} $\eta$ on the brane and
not to physical time. It implies that the energy of KK particles on
a moving AdS brane redshifts like that of massless
particles. 
From this alone one would expect the energy density of 
KK-modes on the brane to decay like $1/a^4$ 
(see also Appendix D of \cite{Gorbunov:2001ge}).

Now, let us define the  normalized ``wave function'' for a graviton
\begin{equation}
\Psi_\alpha(t,y) = \frac{\phi_\alpha(t,y)}{y^{3/2}}\,,
\qquad
2\,\int_{y_b}^{y_s} dy \Psi_\alpha^2(t,y)=1\,.
\end{equation}
From the expansion of the gravity wave amplitude Eq.~(\ref{e: mode expansion})
and the normalization condition  
it is clear that $\Psi_n^2(t,y)$ gives the probability to find 
a graviton of mass $m_\alpha$ for a given (fixed) time $t$ at position 
$y$ in the ${\ZZ}_2$-symmetric AdS-bulk.

\begin{figure}[ht]
\includegraphics[height=5.6cm]{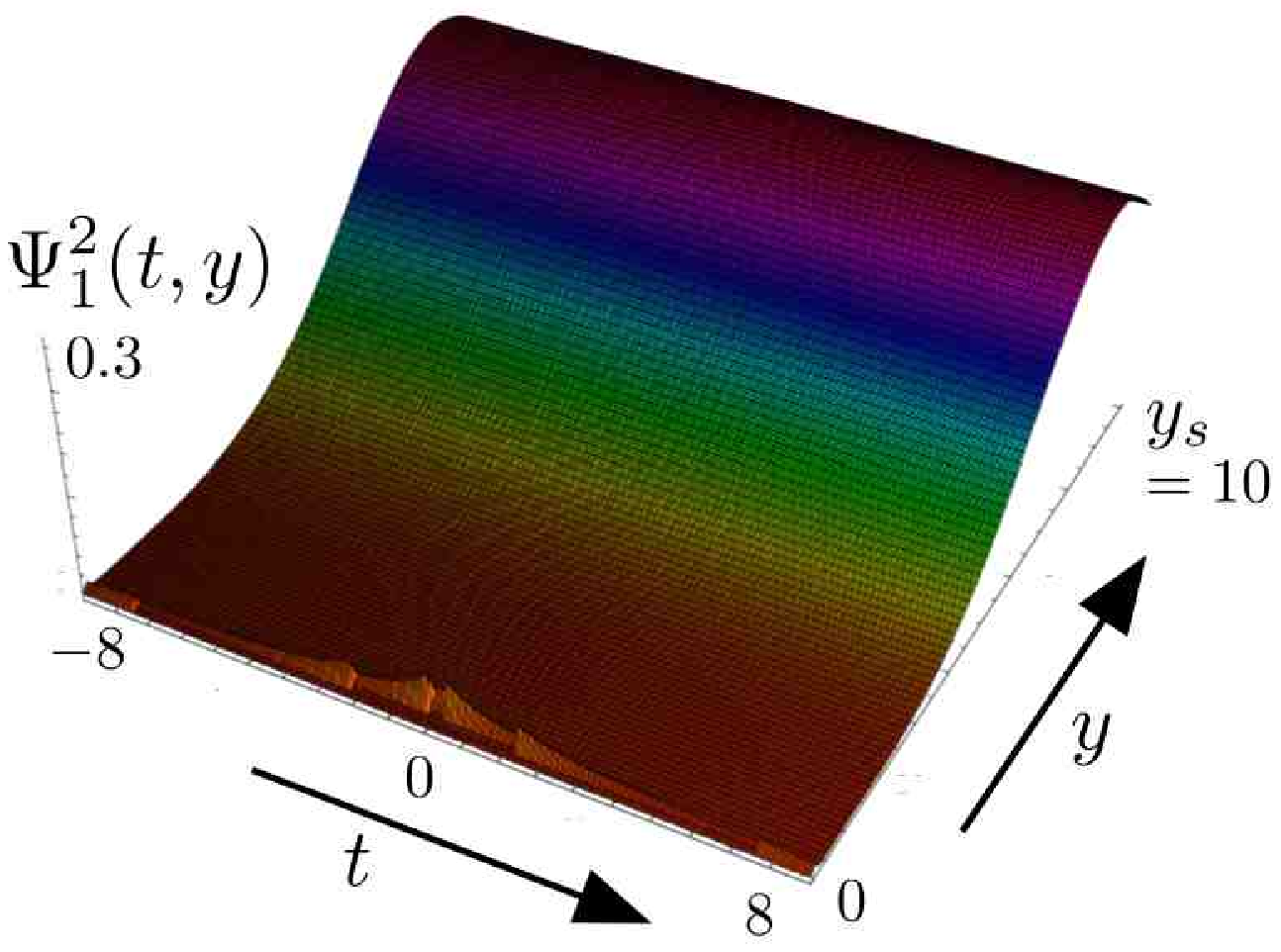} \quad
\includegraphics[height=5.6cm]{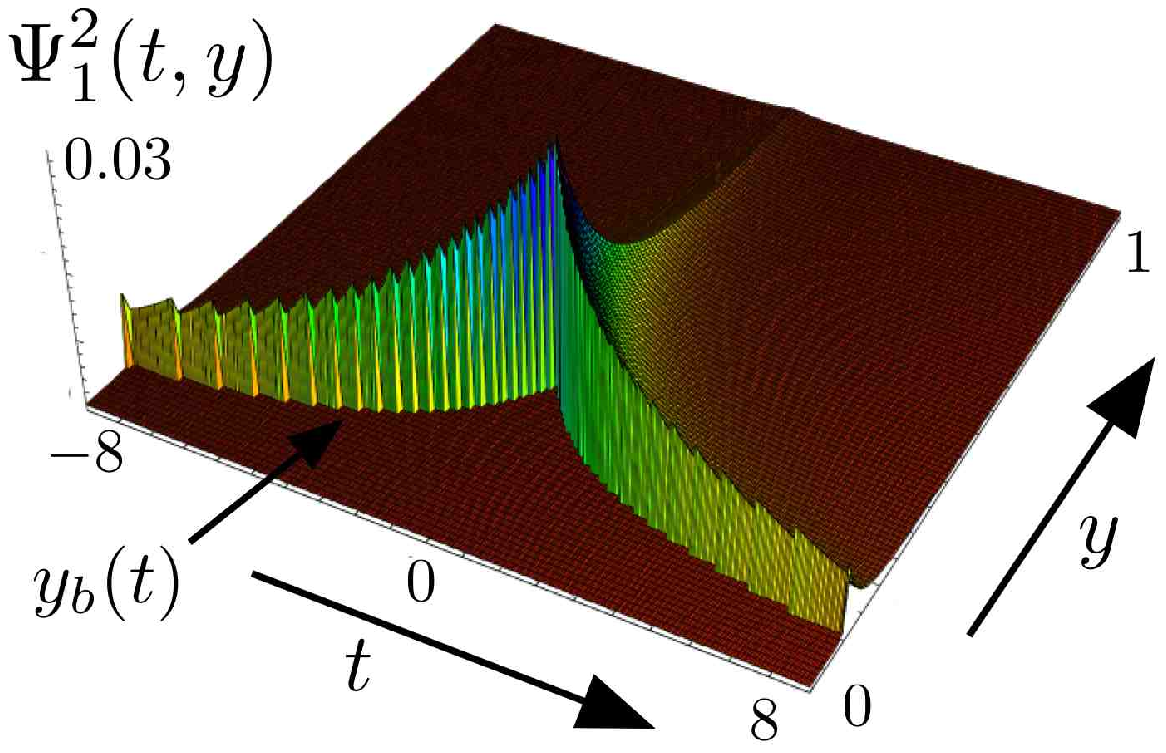}
\caption{Evolution of $\Psi^2_1(t,y) = \phi_1^2(t,y)/y^{3}$ 
corresponding to the probability to find the first KK graviton
at time $t$ at the position $y$ in the AdS-bulk. The static
brane is at $y_s=10L$ and the maximal brane velocity is given by
$v_b=0.1$. On the right hand panel a zoom  into the bulk-region close to 
the moving brane is shown.
\label{f:prob 1}}
\end{figure}

In Fig.~\ref{f:prob 1} we plot the evolution of $\Psi_1^2(t,y)$
under the influence of the brane motion with $v_b= 0.1$. 
For this motion, the physical brane starting at $y_b \rightarrow 0$
for $t \rightarrow -\infty$ moves towards the static 
brane, corresponding to a contracting universe. 
After a bounce, it moves back to the Cauchy horizon, 
i.e. the universe expands.
The second brane is placed at $y_s = 10L$ 
and $y$ ranges from $y_b(t)$ to $y_s$. 
As it is evident from this Figure, $\Psi_1^2$
is effectively localized close to the static brane, i.e. 
the weight of the KK-mode wave function lies
in the region of less warping, far from the physical brane.
Thus the probability to find a KK-mode is larger 
in the region with less warping.
Since the effect of the brane motion on $\Psi_1^2$ is hardly visible
in Fig.~\ref{f:prob 1}, we also show the behavior of $\Psi_1^2$ close to
the physical brane (right hand panel). 

\begin{figure}[ht]
\includegraphics[height=5.6cm]{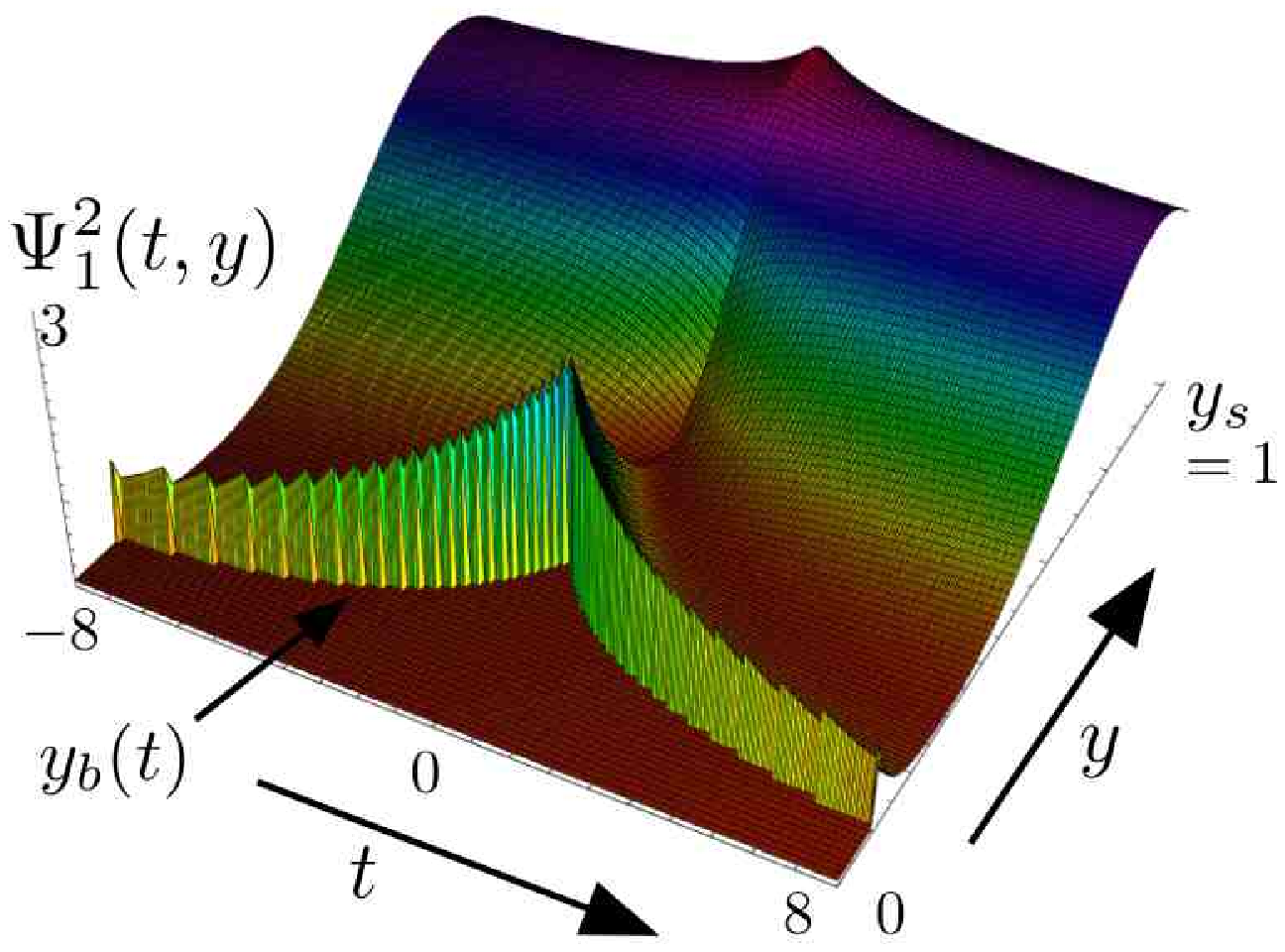} \quad
\includegraphics[height=5.6cm]{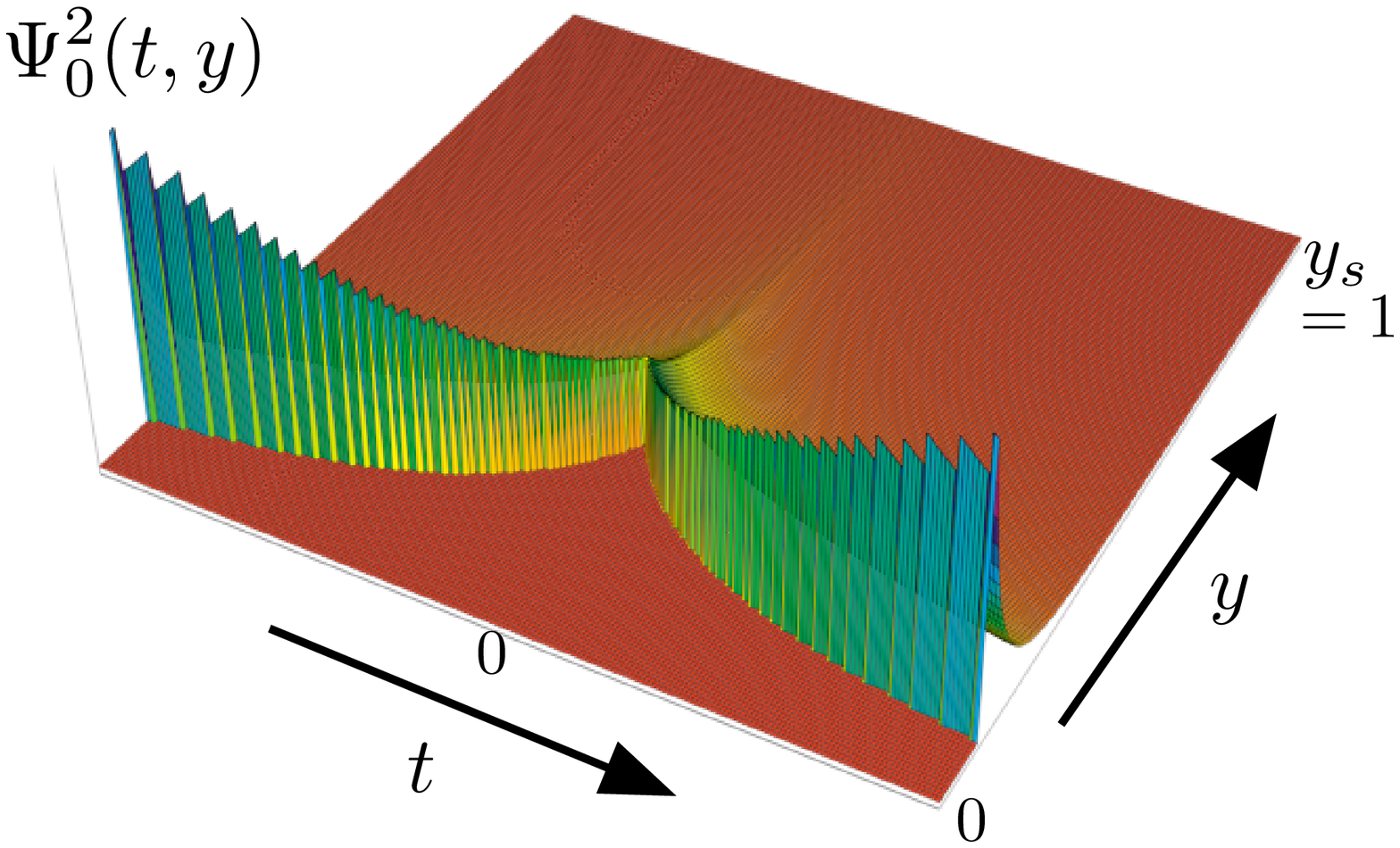}
\caption{Left panel: evolution of $\Psi^2_1(t,y)$ for $y_s=L$ and $v_b=0.1$.
Right panel: localization of four-dimensional gravity on a 
moving brane. Evolution of $\Psi^2_0(t,y)$. Note the opposite behavior 
of zero mode and massive mode. 
\label{f:prob 3}}
\end{figure}

This shows that $\Psi_1^2$ peaks also at the physical 
brane but with an amplitude roughly ten times smaller 
than the amplitude at the static brane. 
While the brane, coming from $t\rightarrow -\infty$, approaches 
the point of closest encounter, $\Psi_1^2$ slightly increases and 
peaks at the bounce $t=0$ where, as we shall see, the production of 
KK particles takes place. 
Afterwards, for $t\rightarrow \infty$, when the brane is
moving back towards the Cauchy horizon, the amplitude
$\Psi_1^2$ decreases again and so does the  probability to find a
KK particle at the position of the physical brane, i.e. in our universe.
The parameter settings used in Fig.~\ref{f:prob 1}
are typical parameters which we use in the numerical simulations.
However, the effect is illustrated much 
better if the second brane is closer to the moving brane. 
In Figure \ref{f:prob 3} (left panel) we show $\Psi_1^2$ for the same parameters 
as in Figure \ref{f:prob 1} but now with $y_s=L$. 
In this case, the probability to find a KK particle on the physical brane
is of the same order as in the region close to the second brane
during times close to the bounce.
However, as the universe expands, $\Psi_1^2$ rapidly decreases
at the position of the physical brane.

The behavior of the KK-mode wave function suggests 
the following interpretation:
If KK gravitons are created on the brane, or equivalently 
in our universe, they escape from the brane into the bulk 
as the brane moves back to the Cauchy horizon, i.e.
when the universe undergoes expansion. 
This is the reason why the power spectrum and the energy density 
imprinted by the KK-modes on the brane decrease faster with the 
expansion of the universe than for the massless zero mode. 

The zero mode, on the other hand, is localized at the 
position of the moving brane. 
The profile of $\phi_0$ does not depend on the extra 
dimension, but the zero-mode wave function $\Psi_0$ does.
Its square is  
\begin{equation}
\Psi_0^2(t,y) = \frac{y_s^2 y_b^2}{y_s^2 - y_b^2}\frac{1}{y^3}
\rightarrow \frac{y_b^2}{y^3} = \left(\frac{L}{a}\right)^2\frac{1}{y^3}
\;\;{\rm if}\;\;y_s \gg y_b~,
\label{e:zero mode in bulk}
\end{equation} 
such that on the brane ($y=y_b)$ it behaves as 
\begin{equation}
\Psi_0^2(t,y_b) \simeq \frac{a}{L}.
\label{e:zero mode on brane}
\end{equation}
Equation (\ref{e:zero mode in bulk}) shows that, 
at any time, the zero mode is localized 
at the position of the moving brane. 
For a better illustration we show 
Eq.~(\ref{e:zero mode in bulk}) in Fig.~\ref{f:prob 3}, right panel 
for the same parameters as in the left panel.
This is the ``dynamical analog'' of the localization mechanism 
for four-dimensional gravity discussed in \cite{RS1,RS2}.

This result is in contradiction with the findings of Ref.~\cite{lang} where 
the authors conclude that for an  observer on the brane KK gravitons behave 
like dust with a negative energy density. To
arrive at this result, they use  Gaussian normal coordinates,
\bea
ds^2 &=& -N^2(t,z)dt^2 +Q^2(t,z)a^2(t)\delta_{ij}dx^idx^j +dz^2 \qquad \mbox{ with}\\
Q  &=& \cosh(z/L) -\ga^{-1}\sinh(|z|/L) \quad N  = 
  \cosh(z/L) -\left(\ga^{-1} -\frac{\dot\ga}{\ga^2H}\right)\sinh(|z|/L) \nonumber \\
  \ga(t)^{-1}&=& \sqrt{(HL)^2 +1} \quad \mbox{ see Eq. (\ref{e:6}).}
\eea
They then argue that at low velocity, $\ga \simeq 1$, one may neglect the difference between $N$
and $Q$ so that one obtains the metric
$$ 
ds^2 \simeq dz^2 + e^{-2|z|/L}\left( -dt^2 + a^2(t)\delta_{ij}dx^idx^j \right)~.
$$
In this metric, the mode equation for the KK modes separates and their time evolution can
be determined by simply solving the time part of the equation, see~\cite{lang}.
There is, however, a flaw in this argument: the above approximation is only valid 
sufficiently close to the brane 
(which is positioned at $z\equiv 0$ in these coordinates), but far from the brane, when, e.g.,
$ (\ga^{-1} -1)\sinh(|z|/L) > \exp(-2|z|/L)$
the above metric is no longer a good approximation and the difference between $N$ and $Q$
does become important. As we have seen, the wave function of the KK gravitons actually is large
far away from the brane and the time dependence enters in an important way in the normalization
of the mode function which changes its scaling with time. 

\subsection{Spectra}
\begin{figure}[ht]
\includegraphics[height=6cm]{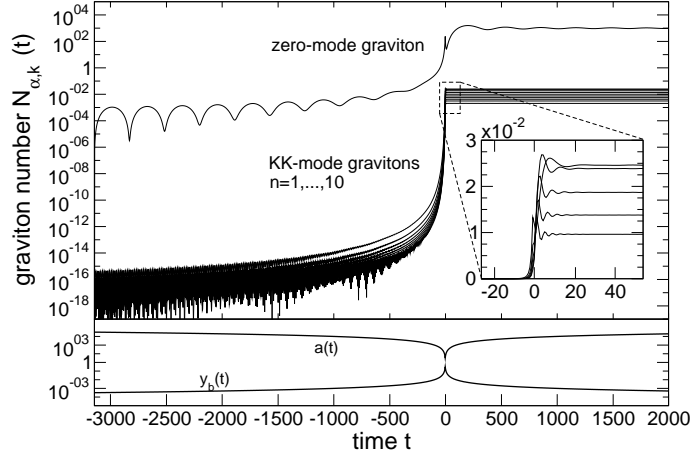}
\caption{Evolution of the graviton number 
${\cal N}_{\alpha,k}(t)$ for the zero mode (one polarization) and the first
ten KK-modes for three-momentum $k=0.01/L$ and
$v_b=0.1$, $y_s=10L$.
\label{genfig1}}
\end{figure}

In Fig.~\ref{genfig1} we show the results of a numerical simulation
for three-momentum $k=0.01/L$, static brane position
$y_s = 10L$ and maximal brane velocity $v_b = 0.1$.
Depicted is the graviton number for one polarization
${\cal N}_{\alpha,k}(t)$ for the zero mode and the first
ten KK-modes as well as the evolution of the scale factor
$a(t)$ and the position of the physical brane $y_b(t)$.

\begin{figure}[ht]
\includegraphics[height=6cm]{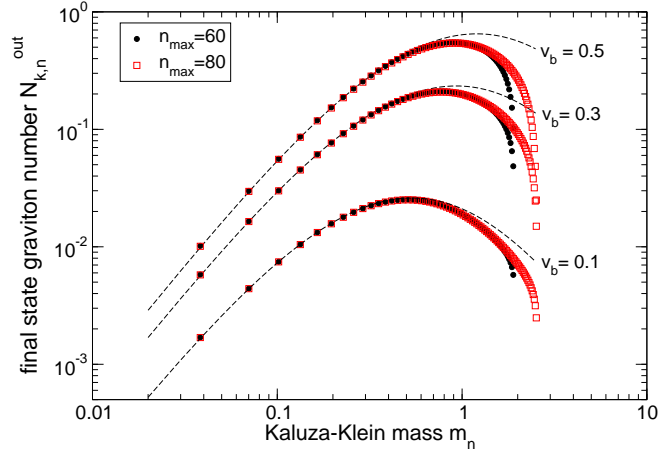}
\caption{Final state KK graviton spectra for $k=0.001$, $y_s=100$, 
  different maximal brane velocities $v_b$ at $t_{\rm out} = 400$ for
  one polarization. The numerical results are compared 
  with the analytical prediction (dashed line).
\label{f:spec}}
\end{figure}

In Fig.~\ref{f:spec} we show some KK spectra which we have obtained by integrating the
equation of motion numerically. More details about the numerics and results for different 
values of the parameters can be found in Ref.~\cite{RD}.
In this paper we also derive an analytical approximation for the spectrum
which is good for KK masses $m_n<1$. The numerical calculations
are in very good agreement with the analytical estimates, where applicable.

Integrating the zero-mode energy density over frequency with a cutoff given
by the strong scale, $k_{\max} =1/L_s$ leads to the following simple result 
for the gravitational wave density parameter~\cite{RD} 
\be
\Omega_{h0} \simeq \frac{v_b}{2}\Omega_\mathrm{rad} \quad \mbox{ so that }
   \quad  v_b \lsim 0.2 \,.
\ee
$\Omega_\mathrm{rad}$ is the density parameter of the relativistic
degrees of freedom at nucleosynthesis, the photon and three species of neutrini.
The limit $v_b<0.2$ follows from the nucleosynthesis constraint which tells us 
that during nucleosynthesis $\Omega_\mathrm{rad}$ should not deviate by more 
than 10\% from its standard value~\cite{book}. The graviton spectrum is blue with
tensor spectral index $n_T=2$. Its amplitude on Hubble scales is therefore
severely suppressed and it leaves no detectable imprint on the cosmic microwave 
background~\cite{book}.

Also the energy density of the KK modes grows like $k^2$ for suffiently large $k$,
$$ \frac{d\rho_{\rm KK}(k)}{d\log k} \propto k^2 ~, \qquad k\gsim 1 $$
and its maximum comes from the cutoff scale  $k_{\max}=1/L_s$. We find
\be
\rho_{\rm KK} \simeq \frac{\pi^5v_b^2}{a^6y_s}\frac{L^2}{L_s^5}\,, \qquad
\left(\frac{\rho_{\rm KK}}{\rho_{\rm rad}}\right)_{\max} 
\simeq 100\,v_b^3 \left(\frac{L}{y_s}\right)\left(\frac{L}{L_s}\right)^2~.
\ee
It is easy to see that low energy requires $y_b<L$ at all times. Therefore, to initiate a
bounce, where $y_b$ should be close to $y_s$, we expect $y_s\lsim L$.
For typical values of the string scale, $L_s\ll L$ and $y_s\sim L$,
 the above ratio is not small and back reaction of the KK gravitons
on the geometry has to be taken into account. The ratio indicated
is the one directly after the big bang. As time goes on the KK mode energy density
dilutes faster than radiation and rapidly becomes subdominant.

\section{Conclusions}
In braneworld cosmology where expansion is mimicked by a brane moving through 
a warped higher dimensional spacetime, the brane motion leads to particle creation 
via the dynamical Casimir effect for all bulk modes. Here we have studied the 
generation of gravitons.

The KK gravitons scale like stiff matter, $\rho_{\rm KK}\propto 1/a^6$, and can therefore 
not represent dark matter. In an
 'ekpyrotic type' scenario with an $\ads$ bulk, 
the nucleosynthesis bound on gravitational waves requires $v_b <0.2$. 
Furthermore, back reaction of KK gravitons on the evolution 
of spacetime is most probably not negligible at early times. 

In the RSII model where only one brane is present, graviton generation 
is negligible~\cite{CDR}. 

\begin{theacknowledgments}
  RD thanks the Organizers of the Spanish Relativity meeting for inviting her to Salamanca,
  to assist and talk at this stimulating meeting. This work is supported by the Swiss National 
  Science Foundation. Thanks go also the the Galileo Galilei Institut in Florence where part 
  of the writing was done. 
\end{theacknowledgments}
\bibliographystyle{aipproc}

\end{document}